\documentclass[twocolumn,showpacs,amssymb,aps,prl,nofootinbib,floatfix,
superscriptaddress]{revtex4-1}

\usepackage{supertabular}
\usepackage{lipsum}
\usepackage{blindtext}
\usepackage{amsfonts}
\usepackage{tensor}
\usepackage{graphicx}
\usepackage{pict2e}
\usepackage{balance}
\usepackage{blindtext}
\usepackage[utf8]{inputenc}

\usepackage{amssymb, amsmath,environ}
\usepackage{comment}

\usepackage{color}
\usepackage[colorlinks=true,urlcolor=black,linkcolor=black,
citecolor=red]{hyperref}


\newcommand{\bse}{\begin{subequations}}
	\newcommand{\ese}{\end{subequations}}
\newcommand{\be}{\begin{equation}}
\newcommand{\ee}{\end{equation}}

\NewEnviron{eqsplit}{%
	\begin{equation}\begin{split}
	\BODY
	\end{split}\end{equation}
}

\makeatletter
\newcommand*\bigcdot{\mathpalette\bigcdot@{.5}}
\newcommand*\bigcdot@[2]{\mathbin{\vcenter{\hbox{\scalebox{#2}{$\m@th#1\bullet$}}}}}
\makeatother

\newcommand{\bea}{\begin{eqnarray}}
\newcommand{\eea}{\end{eqnarray}}
\newcommand{\ba}{\begin{array}}
	\newcommand{\ea}{\end{array}}

\newcommand{\nn}{{\nonumber}}

\def \beq{\begin{equation}}
\def \eeq{\end{equation}}
\def \beqa{\begin{eqnarray}}
\def \eeqa{\end{eqnarray}}

\def \pbar{\bar{p}}

\newcommand{\sNN}{\sqrt{s_{\textrm{NN}}}}

\begin{document}

\title{First flow harmonic of net baryon from directed 
diffusion of stopped baryons}

\author{Tribhuban Parida}
\email[]{tribhu.451@gmail.com}
\affiliation{Department of Physical Sciences,\\ Indian Institute of Science Education and Research Berhampur, Laudigam–760003, Ganjam, Odisha, India}
\affiliation{AGH University of Krakow, Faculty of Physics and Applied Computer Science, aleja Mickiewicza 30, 30-059 Cracow, Poland}

\author{Sandeep Chatterjee}
\email[]{sandeep@iiserbpr.ac.in}
\affiliation{Department of Physical Sciences,\\ Indian Institute of Science Education and Research Berhampur, Laudigam–760003, Ganjam, Odisha, India}

\begin{abstract}
 
We propose a new ansatz for baryon deposition that incorporates salient
features from the baryon junction picture. The deposited baryon is coupled
to a tilted fireball and deployed to study the directed flow of identified 
hadrons in Au+Au collisions at $\sNN=7.7-200$ GeV. We find a parameter
space in the model where the resulting hydrodynamic evolution generates a
suitable flow to describe the directed flow $v_1$ of identified hadrons
including the observed double sign change between $\sNN=7.7-39$ GeV of the
rapidity slopes of the net proton and the net lambda $v_1$. Our model that
includes stopping of baryons and their consequent diffusion within a
relativistic hydrodynamic framework along with a crossover equation of state
as obtained from lattice QCD calculations provides a noncritical baryonic
baseline that is crucial in our ongoing searches for the QCD critical point
and the initial strong electromagnetic field.

\end{abstract}

\maketitle

The transverse phase space of produced hadrons in relativistic heavy 
ion collisions is often studied by performing a Fourier decomposition
\beq
\frac{dN}{d\phi } = N_0\left(1 + 2 v_1\cos(\phi-\psi_\text{RP}) + ...
\right)
\eeq
where $\psi_{\text{RP}}$ is the reaction plane angle in the laboratory 
frame. The first harmonic, $v_1$ is known as the directed flow. There 
have been several model studies of $v_1$ within hydrodynamic 
models~\cite{Ryu:2021lnx,Shen:2020jwv,Bozek:2010bi,Parida:2022lmt,
Jiang:2023fad,Jiang:2021ajc,Jiang:2021foj} including 3-fluid 
model~\cite{Ivanov:2005yw,Ivanov:2014ioa,Konchakovski:2014gda,
Karpenko:2023bok} and transport models~\cite{Reichert:2022qys,
Bleicher:2000sx,Isse:2005nk,Guo:2012qi,Konchakovski:2014gda,Nara:2016hbg,
Zhang:2018wlk,Nara:2020ztb,Nara:2021fuu,Nara:2022kbb,Nara:2021fuu}.
$v_1$ has been demonstrated to be a sensitive probe of the initial 
longitudinal matter deposition~\cite{Bozek:2010bi,Chatterjee:2017ahy} and
QCD equation of state (EoS)~\cite{Ivanov:2014ioa,
Steinheimer:2014pfa,Rischke:1995pe,Nara:2016phs,Hillmann:2018nmd,
Stoecker:2004qu,Brachmann:1999xt,Csernai:1999nf, Nara:2016hbg, 
Steinheimer:2023xzs,OmanaKuttan:2022aml,Steinheimer:2022gqb,
Hillmann:2019wlt,Hillmann:2018nmd,Nara:2017qcg,Reichert:2022gqe}. The mid-rapidity slope of the directed flow, $dv_1/dy$, of baryon exhibits a non-monotonic dependence on the collision energy, which has been proposed as a possible signature of a first-order phase transition between the quark--gluon plasma and the hadronic phase~\cite{Stoecker:2004qu,
Brachmann:1999xt,Csernai:1999nf,Rischke:1995pe,Steinheimer:2014pfa,
Nara:2016phs,Nara:2016hbg,Steinheimer:2022gqb}. Recently, charge 
splitting in $v_1$ has been proposed as a sensitive probe of the 
initial strong electromagnetic (EM) field expected to be generated in 
RHIC with increasing sensitivity as one goes from 
light~\cite{Gursoy:2018yai,Gursoy:2014aka,Nakamura:2022ssn,
Inghirami:2019mkc,STAR:2023jdd,STAR:2023bzh} to 
heavy~\cite{Chatterjee:2017ahy,Sun:2023adv,Das:2016cwd,
Chatterjee:2018lsx} flavors.

While models have been quite successful in describing the charged hadron 
$v_1$~\cite{Bozek:2010bi, Jiang:2021foj, Shen:2020jwv}, $v_1$ of identified hadrons with the splitting in the 
magnitude of $v_1$ between the particle - antiparticle pair has been a 
challenge to capture \cite{Singha:2016mna}. The identified hadrons $v_1$ and in particular the $v_1$ splitting in 
baryon and anti-baryon have been shown to be a sensitive observable of 
the initial baryon deposition scheme~\cite{Du:2022yok, Bozek:2022svy,
Shen:2020jwv}. Currently, there is no first principle understanding of 
the baryon stopping mechanism and the ensuing pre-equilibrium baryon 
dynamics in RHIC. The initial condition in the presence 
of baryons have been either obtained from parametric models 
whereby the thermalised energy-momentum and baryon distribution 
is proposed at a constant proper time hypersurface~\cite{Du:2022yok,
Bozek:2022svy,Shen:2020jwv}, or more microscopic approaches 
whereby the transition from pre-equilibrium dynamics to 
hydrodynamics is implemented locally through dynamical 
rules~\cite{Shen:2023awv,Shen:2022oyg,Shen:2017bsr,Du:2018mpf}.

A natural ansatz for the transverse baryon 
deposition is to consider it to be proportional to the distribution of the 
participant nucleons, carriers of the baryon quantum 
number~\cite{Du:2022yok,Bozek:2022svy,Shen:2020jwv,Denicol:2018wdp}. 
This simple participant scaling of the deposited baryon charge, 
however, substantially overestimates the rapidity odd $v_1$ of 
proton and anti-proton~\cite{Shen:2020jwv}. It has been shown that 
such a participant scaling ansatz along with a tilted distribution 
of the fireball can capture the proton-anti-proton $v_1$ splitting 
at $\sNN=200$ GeV~\cite{Bozek:2022svy}. In addition to the above 
participant deposition, a rapidity even deposition motivated by baryon 
junction model~\cite{Du:2022yok} has been shown to successfully explain the pion 
and proton rapidity odd $v_1$ across  $\sNN=7.7 - 200$ 
GeV. However, a simultaneous description of both 
baryon as well as antibaryon $v_1$ and describing the double sign 
change in net-p and net-$\Lambda$ $v_1$ is yet to be obtained \cite{STAR:2014clz}. 

The baryon junction picture suggests baryon deposition to take place 
via stopping of single as well as double junctions \cite{Kharzeev:1996sq,Vance:1999pr,ToporPop:2004lve,Vance:1998vh}. The transverse profiles 
of these single and double junctions are crucial inputs for phenomenology 
to setup the initial transverse baryon deposition profile. Each nucleon carries 
a baryon junction \cite{Rossi:1977cy,Kharzeev:1996sq,Klein:2026azm,Magdy:2025udq}. The single junction transverse profile is simply given 
by the participant sources transverse profile as obtained in the Glauber model. 
The double junction involves two baryon junctions from two participant nucleons - 
one each from the target and the projectile nucleus. Thus, it is natural to associate the transverse profile of 
the double baryon junctions with the binary collision sources profile 
in the Glauber model \cite{Kharzeev:2000ph}. This results in a novel two component formula for the 
baryon deposition profile at the initial proper time $\tau_0$ receiving contribution 
from both - the participant as well as the binary collision sources terms:
\beqa
  n_B(x,y,\eta_{s}; \tau_0) &=& N_{B} \left[ \left( N_{+}(x,y) f_{+}^{n_B}(\eta_{s}) + N_{-}(x,y) f_{-}^{n_B}(\eta_{s}) \right)\right.\nn\\
                &&\left. \times \left( 1- \omega \right) + N_{bin} (x,y)  F_{bin}^{n_B}(\eta_s) \omega \right] 
 \label{eq:baryon_tilt}
\eeqa
where $N_+$ ($N_-$) are the forward (backward) moving participant sources while $N_{bin}$ is the binary collision source term. Apart from the double junction baryon stopping picture, the $N_{bin}$ term can also be motivated from microscopic transport models, whereby the rapidity loss of a participant baryon source in the pre-equilibrium stage is usually proportional to the number of binary collisions it undergoes~\cite{Jeon:1997bp,De:2022yxq}. $\omega$ controls the relative contribution from the 
participant and the binary collision sources. 

The forward and backward moving participants deposit baryon asymmetrically in rapidity as suggested by the single junction baryon stopping picture \cite{Kharzeev:1996sq}. The rapidity envelope profile for asymmetric baryon depositions are $f_{+}^{n_{B}}(\eta_s)$ and $f_{-}^{n_{B}}(\eta_s)$ which have the same form as in Refs.~\cite{Shen:2020jwv,Denicol:2018wdp}.
Unlike the participant 
sources, the binary collision sources can not be labelled as forward or backward moving and 
hence it is natural to extend them in rapidity ($F_{bin}^{n_B}(\eta_s)$) by a rapidity even profile. 
The double baryon junction stopping picture suggests a plateau baryon deposition profile in rapidity~\cite{Kharzeev:1996sq,Frenklakh:2023pwy,Du:2022yok}. We adopt here the simplest ansatz without introducing any new free parameter:
$F_{bin}^{n_B}(\eta_s) =  f_{+}^{n_{B}}(\eta_s) +  f_{-}^{n_{B}}(\eta_s)$. We have checked that the rapidity odd directed flow is not sensitive to the 
specific choice of the rapidity even profile of the binary collision term. Finally, $N_B$ in Eq.~\ref{eq:baryon_tilt} is 
a normalisation constant to be determined from the condition that the total 
baryon deposited should be equal to the total participants 
$N_{part} = \int \tau_0 d\eta_s dxdy n_B(x,y,\eta_s;\tau_0)$.

\begin{figure}
 \begin{center}
  \includegraphics[scale=0.6]{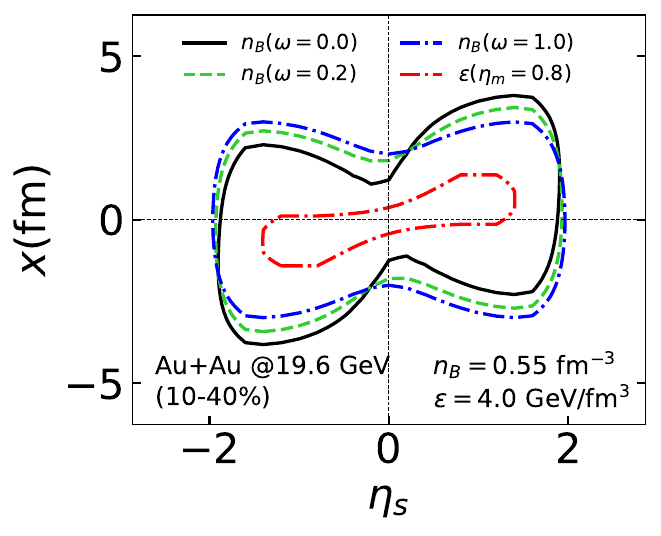}
 \caption{Contours of constant net-baryon density in the reaction plane for different values of the baryon tilt parameter $\omega$ in Au+Au collisions at $\sqrt{s_{\mathrm{NN}}}=19.6$ GeV and 10--40\% centrality. The contour corresponding to the energy density profile obtained using the tilted initial condition (TIC) model with $\eta_m=0.8$ is also shown for comparison, illustrating the relative tilt between the energy density and net-baryon density profiles.}
 \label{fig:contour}
 \end{center}
\end{figure}

The tilted initial condition (TIC) model provides an initial energy density profile, $\epsilon$, that successfully describes the directed flow of charged hadrons~\cite{Bozek:2010bi}. In this model, a free parameter, $\eta_m$, controls the tilt of the initial energy density profile in the reaction plane. In the present study, the initial net-baryon density profile given by Eq.~\ref{eq:baryon_tilt} is coupled to the TIC model. The parameter $\omega$ controls the tilt of the net-baryon density profile. Consequently, the parameters $\eta_m$ and $\omega$ together determine the relative tilt between the energy density and net-baryon density profiles. This effect is illustrated in Fig.~\ref{fig:contour}, which shows contours of constant net-baryon density for different values of $\omega$ in Au+Au collisions at $\sqrt{s_{\mathrm{NN}}}=19.6$ GeV and 10--40\% centrality. For comparison, the contour corresponding to the energy density profile with $\eta_m=0.8$ is also shown.

The resulting initial condition is evolved in several stages: dissipative hydrodynamic evolution including baryon diffusion using the MUSIC code \cite{Schenke:2010nt, Paquet:2015lta, Schenke:2011bn} that evolves the baryon alone while approximating for strangeness and charge from local conditions of strangeness neutrality ($n_S=0$) and a fixed baryon-to-charge density ratio ($n_Q = 0.4, n_B$)~\cite{Denicol:2018wdp}, particlization via the iSS code \cite{Shen:2014vra,https://github.com/chunshen1987/iSS}, and hadronic transport using the UrQMD \cite{Bass:1998ca, Bleicher:1999xi}. In the hydrodynamic phase, we employ the NEoS-BQS equation of state (EoS) \cite{Monnai:2019hkn}, which imposes the above constraints for strangeness and charge. These constraints result in non-zero spacetime dependent strange and electric charge chemical potentials ($\mu_S$ and $\mu_Q$) at finite baryon density. This induces splitting in $v_1$ of particle-antiparticle meson pairs with opposite electric charge and strangeness like $\pi^+-\pi^-$ and $K^+-K^-$ in addition to that between baryon-antibaryon pairs.

All model predictions for observables at a given centrality are obtained from single-shot hydrodynamic simulations using a smooth, event-averaged initial profile corresponding to that centrality class~\cite{Shen:2020jwv, Du:2022yok}. We have performed simulations of Au+Au collisions at $(10-40)\%$ centrality across a range of RHIC Beam Energy Scan (BES) collision energies, from $\sqrt{s_{NN}} = 7.7$ to 200 GeV. The hydrodynamic evolution begins at a constant proper time $\tau_0$ that ranges between 0.6 fm (at $\sNN=$ 200 GeV) to 3.6 fm (at $\sNN=$ 7.7 GeV), chosen to be greater than the nuclei passage time~\cite{Shen:2017bsr}.  We use a constant shear viscosity to entropy density ratio, $\eta/s = 0.08$, across all $\sNN$. For baryon diffusion, we adopt a temperature and baryon chemical potential dependent diffusion coefficient derived from kinetic theory as in Ref. \cite{Denicol:2018wdp}.
Our model parameters were chosen to simultaneously reproduce the rapidity dependence of charged particle yields, net-proton yields, and the directed flow $v_1(y)$ of $\pi^+$, protons, and anti-protons. Notably, the model parameters, including the tilt parameters, were tuned independently at each collision energy to achieve the best possible agreement with the corresponding experimental data. A thorough, quantitative extraction of parameter values and study of their interdependencies would require a dedicated Bayesian analysis, which we have not performed in this initial study but plan to pursue in the future.

\begin{figure}
 \begin{center}
  \includegraphics[scale=0.6]{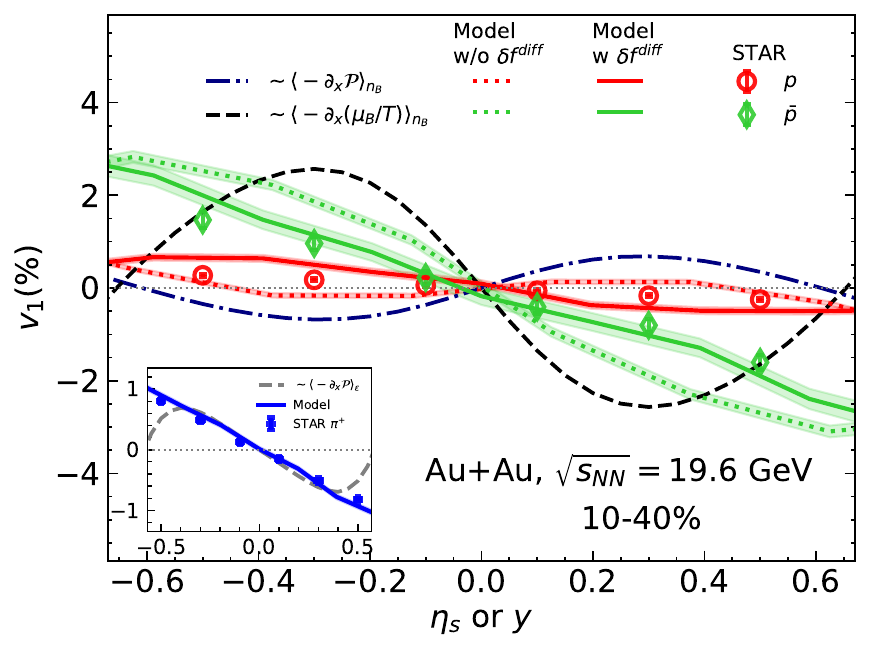}
 \caption{
 (Color online) Directed flow, $v_1$, as a function of rapidity for protons (red), anti-protons (green), and $\pi^+$ mesons (blue, inset) in Au+Au collisions at $\sqrt{s_{NN}}=19.6$ GeV and 10--40\% centrality. The proton and anti-proton results are shown both with (solid lines) and without (dotted lines) the contribution from baryon diffusion. The rapidity dependence of the estimators $\langle -\partial_x \mathcal{P} \rangle_{n_B}$ (dashed) and $\langle -\partial_x (\mu_B/T) \rangle_{n_B}$ (dash-dotted) is also shown, while $\langle -\partial_x \mathcal{P} \rangle_{\epsilon}$ is displayed in the inset for comparison with the $\pi^+$ results. The symbols represent measurements by the STAR Collaboration~\cite{STAR:2014clz}. }
 \label{fig:delatf_diff_v1}
 \end{center}
\end{figure}

In Fig.~\ref{fig:delatf_diff_v1}, we illustrate the role of baryon diffusion on pressure gradient in determining the final splitting of $v_1$ in baryon anti-baryon pair by focusing on rapidity differential $v_1$ of $p$ and $\pbar$ at $\sqrt{s_{NN}} = 19.6$ GeV. The presence of a dissipative baryon diffusion current during the hydrodynamic stage shows up in the final Cooper-Frye particalisation via $\delta f^{\text{diff}}$, the  non-equilibrium contribution due to baryon diffusion on the hadron phase space distribution~\cite{Denicol:2018wdp}. The model results with and without $\delta f^{\text{diff}}$ clearly demonstrates the significant effect of baryon diffusion on $v_1$ of $p$ and $\pbar$. 

$v_1$ signifies the asymmetric flow of matter along and opposite to the impact parameter axis, conventionally chosen as the $+x$ direction. This flow asymmetry originates from asymmetric pressure gradient in the initial state. Consequently, the development of the net transverse flow and the ultimate sign of $v_1$ in the final state can be inferred from the estimator $\langle -\partial_x \mathcal{P} \rangle_\epsilon$ \cite{Bozek:2010bi,Jiang:2021ajc}. In a baryon-free medium, this estimator alone is sufficient to characterize directed flow. In the presence of finite baryon density, however, baryonic currents also play a role. The net-baryon current, $J_B^\mu = n_B u^\mu + q_B^\mu$, contains both equilibrium transport (advection with the fluid velocity) and a diffusion component driven by spatial gradients of $(\mu_B/T)$. These effects are quantified through two additional estimators: $\langle -\partial_x \mathcal{P} \rangle_{n_B}$, describing baryon transport along the matter flow direction, and $\langle -\partial_x (\mu_B/T) \rangle_{n_B}$, capturing baryon diffusion induced dissipative contributions. All these estimators are defined compactly as
\beq
\Big\langle -\partial_x \Phi \Big\rangle_{w}(\eta_s) \equiv
\frac{\int d^2 r_\perp  \big(-\partial_x \Phi(x,y,\eta_s)\big) w(x,y,\eta_s)}
{\int d^2 r_\perp  w(x,y,\eta_s)} ,
\label{eq:estimator_2}
\eeq
with $\Phi \in \{\mathcal{P}, \mu_B/T\}$ and $w \in \{\epsilon, n_B\}$.
In particular, for proton, in the absence of $\delta f^{\text{diff}}$, the flow is baryon advection along the bulk and follows  $\langle -\partial_x \mathcal{P} \rangle_{n_B}$, while in the presence of $\delta f^{\text{diff}}$, the flow is diffusion dominated and follows $\langle -\partial_x (\mu_B/T) \rangle_{n_B}$ with a change in sign in the rapidity slope. For completeness, $v_1$ of $\pi^+$ is shown in the inset. The model results 
follow $\langle -\partial_x \mathcal{P} \rangle_{\epsilon}$ as expected. Having understood the role of energy and baryon deposition and 
consequent diffusion in describing the identified hadron $v_1$ at $\sNN=19.6$ GeV, we will now shift our focus to the $\sNN$ dependence 
of $v_1$.

\begin{figure}
 \begin{center}
  \includegraphics[scale=0.5]{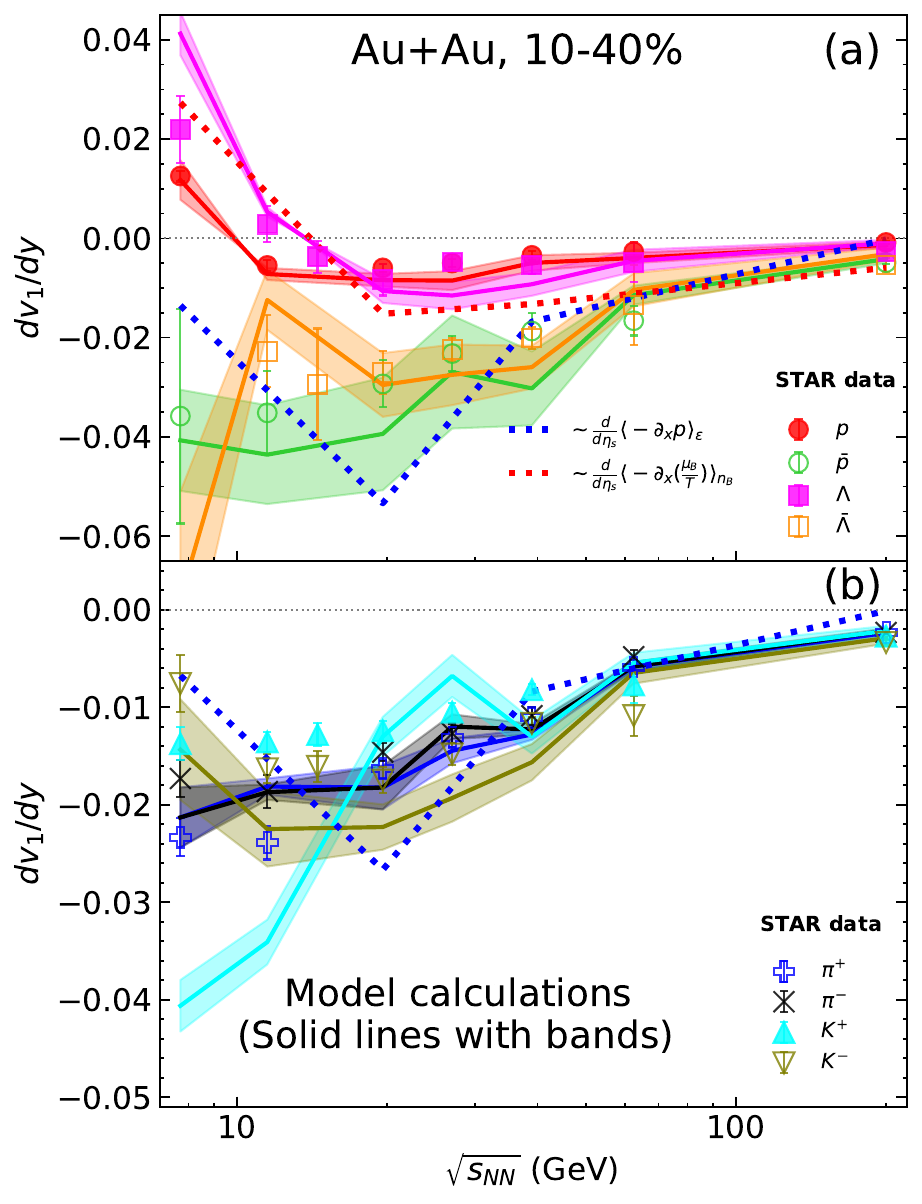}
 \caption{(Color online) Collision-energy dependence of the mid-rapidity slope of the directed flow, $dv_1/dy$, for identified hadrons in Au+Au collisions at 10--40\% centrality. Panel (a) shows the results for baryons and anti-baryons, while panel (b) presents the corresponding results for mesons. The model calculations are shown as solid lines, and the measurements by the STAR Collaboration are represented by symbols~\cite{STAR:2014clz,STAR:2017okv}. The space--time rapidity slopes of the initial-state estimators, $\langle -\partial_x \mathcal{P} \rangle_{\epsilon}$ and $\langle -\partial_x (\mu_B/T) \rangle_{n_B}$, are also shown as dotted lines to guide the interpretation of the model results.}
 \label{fig:dv1dy_id}
 \end{center}
\end{figure}

The collision-energy dependence of the mid-rapidity slope of the directed flow, $dv_1/dy$, for identified hadrons is shown in Fig.~\ref{fig:dv1dy_id}. Panel (a) presents the results for baryons and anti-baryons, while panel (b) shows those for mesons. The same kinematic acceptance and fitting procedure as used in the experimental analyses are adopted to extract $dv_1/dy$ from the model calculations~\cite{STAR:2014clz,STAR:2017okv}. As the collision energy decreases, baryon stopping becomes increasingly significant, leading to a substantial increase in the average value of $\mu_B/T$ around mid-rapidity. This is accompanied by an enhanced splitting in $v_1$ between baryons and anti-baryons, indicating increasingly strong spatial gradients in the initially stopped net-baryon distribution. In contrast, because $\mu_{S,Q}/T \ll \mu_B/T$, the splitting between particle--antiparticle pairs in the meson sector, which differ only in strangeness and/or electric charge, remains considerably smaller. The measured directed flow of $\pi^+$, $p$, and $\bar{p}$ is used to constrain the tilts of the initial energy density and net-baryon density profiles in the reaction plane, while the remaining model parameters are fixed independently from multiplicity measurements. The directed flow of all other identified hadrons is therefore a genuine prediction of the model, obtained without any additional parameter tuning.

While the agreement for mesons carrying strangeness and/or electric charge is qualitative and gradually deteriorates with decreasing $\sqrt{s_{NN}}$, as $\mu_{S,Q}/T$ increases, the model provides good quantitative agreement for the strange baryons $\Lambda$ and $\bar{\Lambda}$, whose directed flow is predominantly governed by the baryon current. The model results closely follow the initial-state indicators discussed in Fig.~\ref{fig:delatf_diff_v1}. Specifically, the directed flow of mesons and anti-baryons follows the space--time rapidity slope of $\langle -\partial_x \mathcal{P} \rangle_{\epsilon}$, whereas that of baryons tracks $\langle -\partial_x (\mu_B/T) \rangle_{n_B}$, indicating that baryon diffusion dominates over baryon advection in determining the baryon directed flow.

The quantitative description of the baryon and anti-baryon $v_1$ demonstrates that the model successfully captures the advection and diffusion of net baryon number. In contrast, the dynamics of the other conserved QCD charges, namely strangeness and electric charge, are not evolved independently, and their associated dissipative currents are not included. Consequently, the agreement with data for hadrons carrying these charges becomes progressively worse at lower collision energies, where larger gradients in the corresponding chemical potentials are expected to enhance dissipative effects.

\begin{figure}
 \begin{center}
  \includegraphics[scale=0.5]{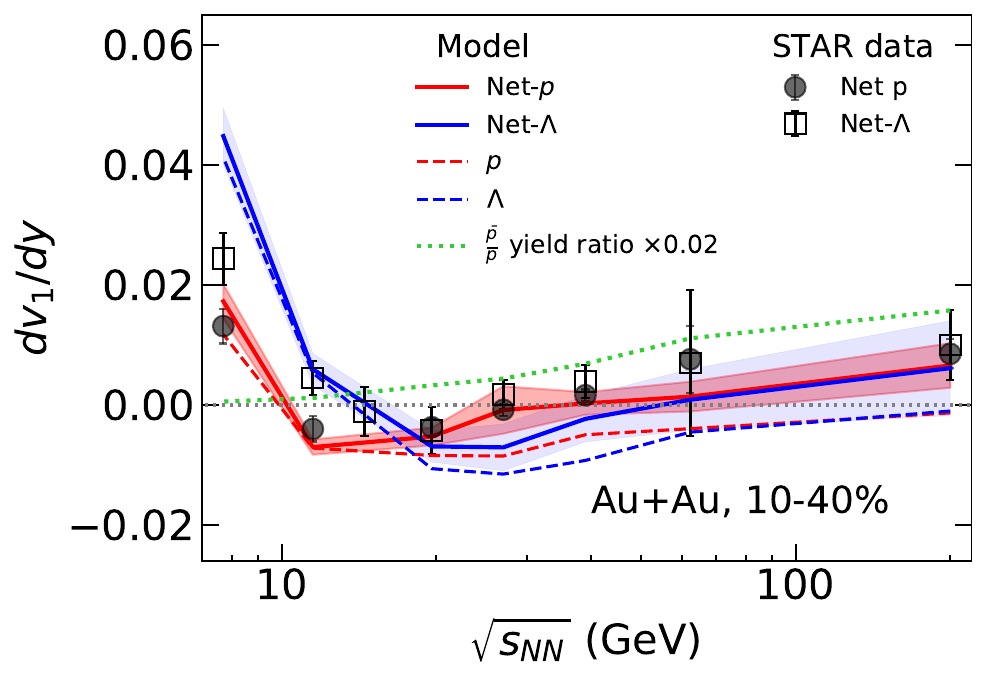}
 \caption{(Color online) Beam energy dependence of the directed flow slope ($dv_1/dy$) for net-protons and net-$\Lambda$, as defined in Eq. \eqref{eq:netXv1} for Au+Au collisions of 10–40\% centrality. Model calculations (lines with shaded bands) are compared with experimental measurements from the STAR Collaboration~\cite{STAR:2014clz,STAR:2017okv}. The shaded bands represent statistical uncertainties in the model calculations. For reference, the model results for $dv_1/dy$ of protons and $\Lambda$ are also shown, but error bars are omitted for these species. }
 \label{fig:dv1_dy_nep_netl}
 \end{center}
\end{figure}

The difference in the $v_1$ of baryon - 
antibaryon pair with decreasing $\sNN$ is showcased by the net-proton and 
net-$\Lambda$ $v_1$ measured by the STAR collaboration~\cite{STAR:2014clz,
STAR:2017okv}
\beq
 [v_1(y)]_{Net-X} =  \frac{\frac{dN^{X}}{dy}[v_1(y)]_{X} - 
 \frac{dN^{\bar{X}}}{dy} [v_1(y)]_{\bar{X}} } {\frac{dN^{X-\bar{X}}}{dy}}
 \label{eq:netXv1}
\eeq
This observable shows an intriguing double sign change structure for both, 
$X=$ p and $\Lambda$ between $\sNN=7.7-39$ GeV which has been a challenge 
to capture in models. At low energies, as anti-baryon yield is negligible 
compared to baryon, Eq.~\ref{eq:netXv1} suggests that 
the net-baryon $v_1$ effectively becomes equal to that of the baryon alone. However, with increasing 
$\sNN$, the anti-baryon yield increases and starts 
contributing to the net-baryon $v_1$. As clear from its definition, a good description of yield as well as $v_1$ 
of baryon and anti-baryon simultaneously is needed in order to describe the net-baryon $v_1$.

Fig. \ref{fig:dv1_dy_nep_netl} shows the collision energy dependence of the mid-rapidity slope of $v_1$ for net-$p$ and net-$\Lambda$ from our model, alongside experimental data \cite{STAR:2014clz,STAR:2017okv}. The model reproduces the experimentally observed double sign change in $dv_1/dy$ for both net-protons and net-$\Lambda$ baryons. For comparison, we also show the $dv_1/dy$ values for $p$ and $\Lambda$. At the lower $\sqrt{s_{NN}}$, the mid-rapidity yiel ratio of $\pbar/p$ is negligible and therefore $dv_1/dy$ of net-$p$ is similar to $p$. As $\pbar/p$ rises with increasing $\sqrt{s_{NN}}$, the slopes of net-$p$ and $p$ differ significantly with opposite signs. Thus, to describe the full beam energy dependence of net-proton $dv_1/dy$, the model must simultaneously capture both the yield and the directed flow of proton as well as anti-proton. A similar trend observed for net-$\Lambda$ is also captured by our model.

In hydrodynamic simulations at finite baryon density, an essential input is the initial condition for the net-baryon distribution, which is directly linked to the physics of baryon stopping. In this work, we have proposed a novel ansatz for the initial baryon distribution. When used as input in a hybrid framework—combining hydrodynamic evolution with hadronic transport—we successfully describe the directed flow ($v_1$) of identified hadrons, including $\pi^\pm$, $K^\pm$, $p$, $\bar{p}$, $\Lambda$, and $\bar{\Lambda}$. In particular, our model captures the splitting of $v_1$ between baryons and anti-baryons, a long-standing challenge for models to capture due to improper modelling of initial baryon stopping.

Unlike previous approaches that assumed the transverse net-baryon distribution to be proportional solely to participant density, our Glauber-geometry-based model introduces a two-component baryon stopping scheme. In this framework, the net-baryon profile receives contributions from both participant and binary collision sources. This approach can be interpreted within the baryon junction picture, where single-junction stopping corresponds to participants sources and double-junction stopping are associated with binary collision sources.

Our model generates a tilted net-baryon distribution in the reaction plane, where the degree of tilt is governed by a free parameter that controls the relative contributions of baryon stopping from participants and binary collisions. This relative tilt between the energy and baryon density profiles plays a crucial role in determining the net baryonic flow and, ultimately, the sign of the directed flow ($v_1$) of baryons. We have investigated in detail the underlying mechanism responsible for the generation of asymmetric baryon flow that leads to the observed $v_1$ of both baryons and anti-baryons. Specifically, we have demonstrated how the relative tilt between the energy and net-baryon distributions creates a dipolar asymmetry in the baryon profile with respect to the pressure distribution at non-zero space-time rapidity. This asymmetry contributes to the equilibrium part of the baryonic flow by dictating the amount of baryon number that is advected along the direction of energy flow. Additionally, the resulting asymmetric profile of $\mu_B/T$ induces a directed diffusion current. The competition between advection and diffusion of baryons at non-zero rapidities gives rise to the net directed flow observed in the final state.

In our model, we successfully reproduce the sign change of the mid-rapidity slope of $v_1(y)$, $dv_1/dy$ of proton at lower collision energies—a feature that has previously been suggested as a possible signature of a first-order phase transition. Notably, our simulations employ a crossover equation of state (EoS), indicating that the observed sign change can instead be attributed to the significant influence of baryon diffusion at low collision energies. Furthermore, our model captures the double sign change in the $dv_1/dy$ of both net-protons and net-$\Lambda$ baryons within $\sqrt{s_{NN}} = 7.7-39$ GeV. Given its success in describing these experimental observations, our model provides a valuable framework for further exploring the dynamics of baryons in relativistic heavy-ion collisions. In particular, it offers a promising avenue for constraining fundamental transport coefficients such as the baryon diffusion coefficient. Additionally, it can serve as a robust baseline for interpreting future experimental results related to the search for the QCD critical point and electromagnetic field-induced signals at RHIC Beam Energy Scan (BES) energies.

\section{Acknowledgement}
TP acknowledges support from the Polish Ministry of
Science and Higher Education and from the Polish National
Science Centre Grant No. 2023/51/B/ST2/01625.

\bibliography{v1_ref}

\end{document}